\documentstyle[12pt]{article}
\newcommand{\dfrac}[2]{{\displaystyle{\frac{#1}{#2}}}}

\newcommand{\journal}[4]{{\em #1~}{\bf #2}\,(19#3)\,#4;}

\newcommand{\cmp}{\journal {Comm. Math. Phys.}}

\newcommand{\np}{\journal {Nucl. Phys.}}
\newcommand{\pl}{\journal {Phys. Lett.}}

\newcommand{\G}{\Gamma}
\newcommand{\D}{\Delta}
\renewcommand{\a}{\alpha}
\renewcommand{\b}{\beta}
\renewcommand{\d}{\delta}
\newcommand{\e}{\varepsilon}

\newcommand{\g}{\gamma}
\newcommand{\k}{\kappa}
\newcommand{\x}{\xi}
\renewcommand{\l}{\lambda} \renewcommand{\L}{\Lambda}
\newcommand{\m}{\mu}
\newcommand{\n}{\nu}
\newcommand{\mn}{{\mu\nu}}

\renewcommand{\o}{\omega} \renewcommand{\O}{\Omega}

\renewcommand{\r}{\rho}
\newcommand{\s}{\sigma} \renewcommand{\S}{\Sigma}
\newcommand{\th}{\theta}
\renewcommand{\t}{\tau}

\renewcommand{\AA}{{\cal A}}

\newcommand{\LL}{{\cal L}}
\newcommand{\MM}{{\cal M}}
\newcommand{\NN}{{\cal N}}

\newcommand{\SS}{{\cal S}}

\newcommand{\WW}{{\cal W}}

\newcommand{\ZZ}{{\cal Z}}

\newcommand{\complex}{{\kern .1em {\raise .47ex
\hbox {$\scriptscriptstyle |$}}
    \kern -.4em {\rm C}}}
\newcommand{\real}{{{\rm I} \kern -.19em {\rm R}}}
\newcommand{\rational}{{\kern .1em {\raise .47ex
\hbox{$\scripscriptstyle |$}}
    \kern -.35em {\rm Q}}}

\newcommand{\pa}{\partial}

\newcommand{\ie}{{{\em i.e.}\ }}

\newcommand{\sla}{\raise.15ex\hbox{$/$}\kern -.57em}

\newcommand{\twiddle}{\lower.9ex\rlap{$\kern -.1em\scriptstyle\sim$}}

\newcommand{\dxm}{dx^\mu }
\newcommand{\dxn}{dx^\nu }

\newcommand{\vf}{{\varphi}}
\newcommand{\pam}{{\partial_\mu}}
\newcommand{\pan}{{\partial_\nu}}
\newcommand{\pal}{{\partial_\l}}

\newcommand{\Ot}{{\tilde\Omega}}
\newcommand{\eq}{\begin{equation}}
\newcommand{\eqn}[1]{\label{#1}\end{equation}}
\newcommand{\eea}{\end{eqnarray}}
\newcommand{\eqa}{\begin{eqnarray}}
\newcommand{\eqan}[1]{\label{#1}\end{eqnarray}}
\newcommand{\ba}{\begin{array}}
\newcommand{\ea}{\end{array}}
\newcommand{\eqac}{\begin{equation}\begin{array}{rcl}}
\newcommand{\eqacn}[1]{\end{array}\label{#1}\end{equation}}

\def\non{\nonumber\\}

\def\6{\partial}
\def\={\!\!\!&=&\!\!\!}
\def\+{\!\!\!&&\!\!\!+~}
\def\-{\!\!\!&&\!\!\!-~}

\def\ve{\varepsilon}
\renewcommand{\title}[1]{\null\vspace{25mm}

\noindent{\Large{\bf #1}}\vspace{10mm}

\noindent {\large  }}
\newcommand{\authors}[1]{\noindent{\large #1}\vspace{3mm}}

\newcommand{\address}[1]{\noindent #1\vspace{5mm}

}
\renewcommand{\abstract}[1]{\vspace{10mm}

\noindent{\small{\em Abstract.} #1}\vspace{2mm}

} 
\begin{document}

\setcounter{page}{1}
\thispagestyle{empty}\hspace*{\fill}REF. TUW 97-16
\begin{center}
\title{Ultraviolet and Infrared Finiteness in Two\\
       Dimensional Curved Space--Time}
\authors{H. Zerrouki\footnote{Work supported in part by the
``Fonds zur F\"orderung der Wissenschaftlichen Forschung'',
under Contract Grant Number P11582 -- PHY.}}%\\
%\end{center}

%\begin{center}
\address{Institut f\"ur Theoretische Physik,
         Technische Universit\"at Wien,\\
         Wiedner Hauptstra\ss e 8-10,
         A-1040 Wien, Austria} %\\
\end{center}

\abstract{

Different models of field theories in two dimensions can be described by  
the action $Tr\int \vf F$. In the presence of a curved background,
we construct a local supersymmetry--like transformations under which
the action is invariant. Furthermore,
by analysing the cohomology of the theory we show the absence of
anomalies. Also the ultraviolet as well as the infrared 
finiteness of the theory are proven at all orders of 
perturbation theory. 
}

%\centerline{}
%\centerline{}
%\leftline{Aug. 1997}
%\centerline{}
%\centerline{}

\newpage

\section{Introduction}

Two dimensional fields models \cite{blau93} play an important role 
in physics as
well as in mathematics. For instance, 
conformal field theories and sigma models lead
to the development of string theory. On the other hand, by studying 
low dimensional theories one can gain experience which could help
to tackle more complicated problems already present in the 
4--dimensional world.\\
2D physics has some interesting features, like for instance the action
describing the Yang--Mills model with vanishing coupling constant 
and the action of the Jackiw--Teitelboim model for 2D gravity have 
the same form, namely $Tr \int \vf F$. The same 
action is obtained by compactifying the Chern--Simons model 
\cite{witcs} on a circle.
This was studied by many authors \cite{2dgravity}, \cite{2dgauge} 
and \cite{blasi}.
In this work we will generalize the analysis of \cite{blasi} to be valid
in a curved space--time.     
The main ingredient of our analysis is the use of the 
Landau gauge where local supersymmetry--like transformations are manifest
\cite{ms}, \cite{mwo}.
Another interesting property of the Landau gauge is the existence of
the ghost equation \cite{bps} 
which is very helpful for the analysis.
Our strategy is to follow an algebraic way \cite{olivier2}
and show the stability of the theory. Furthermore, the calculation of 
possible counterterms is carried out essentially by using cohomology 
techniques \cite{dix}, \cite{brandt}, 
\cite{sor}.
%%%%%%%%%%%%%%%%%%%%%%%%%%%%%%%
The advantage in using the framework of algebraic renormalization
is that one does not need to specify a subtraction scheme like
for instance the BPHZ or the dimensional regularization schemes.
On the other hand, such a scheme should exist, a fact which will
allow us to use the above mentioned algebraic renormalization.
This fact limited our quantum analysis to be valid only in curved,
topologically trivial, and asymptotically flat manifolds.\\
This work is organized as follows: 
in section $2$ we present the classical analysis of the two
dimensional model considered in a 
curved space--time. 
We begin by describing the classical theory, and we show that local 
susy--like transformations do exist.
This section is also devoted to the discussion of the infrared 
regularization of the ghost--antighost propagator and its implications. 
Then, we generalize the local susy--like transformations of
the model and we show that the corresponding Ward identity is 
linearly broken. \\
Next, in section 3, we construct the most general counterterm which is, 
in turn, forbidden by the ghost equation. Therefore no deformations are 
allowed and the theory is finite. \\
In section 4, the $2$D theory is proved to be anomaly free. This last 
result allows us to extend the classical analysis to all 
orders of perturbation theory.
Hence the finiteness of the two dimensional model, defined on a curved 
manifold, is proven at all orders of perturbation theory.

%******************************************************************

\newpage

\section{The model and its infrared regularization}

We devote this section to the classical analysis and the infrared
regularization of the $2D$ theory, considered on a two dimensional 
curved manifold $\MM$, in the Landau gauge.\\ 
The most important feature of such a field model is that
it is topological \cite{tft} and possesses the following invariant 
action, which is metric\footnote{Here are our conventions: we denote 
the space--time metric by $g_\mn$, its inverse by $g^\mn$ and its 
determinant by $g$. The scalar density $\sqrt g$ has weight +1, whereas 
the volume element density $d^2x$ has weight -1,
which means that the {\it invariant} volume element $dV=\sqrt g d^2x$ 
has vanishing weight, then transforms as a scalar under diffeomorphisms.}  
independent:
\begin{equation}  
\label{inv}
\Sigma_{inv} = \frac{1}{2} \int_{\cal M} d^{2} x
               \varepsilon^{\mu\nu} F^{a}_{\mu\nu}
               \varphi^{a},
\end{equation} 
where $ \varepsilon^{\mu\nu} $ is a second rank antisymmetric
tensor\footnote{For the Levi--Civita density $\ve^\mn$ we choose
$\ve^{12}=1$. Furthermore, we have
$\ve_{\a\b}=g^{-1}g_{\m\a}g_{\n\b}\ve^\mn$, where $\ve^\mn$ has weight
+1 and $\ve_\mn$ has weight -1.}
density of weight 1, and the field strength
$ F^{a}_{\mu\nu} $ is given
by:
\begin{equation}
\label{curv}
F^{a}_{\mu\nu} = \partial_{\mu} A^{a}_{\nu} -
                 \partial_{\nu} A^{a}_{\mu} +
                 f^{abc} A^{b}_{\mu} A^{c}_{\nu}.
\end{equation}
Here $ f^{abc} $ denotes the structure constants of the
gauge group, which is supposed to be a compact Lie
group and all the fields belong to its adjoint 
representation. $A^{a}_{\mu}$ is the gauge field. \\
%%%%%%%%%%%%%%%%%%%%%%%%%%%%%%%%%%%%%%%%%%%%%%%%%%%%%%%%%%%%%%%%%%%%%%
Now, we briefly review some connections of the 
action (\ref{inv}) with different two dimensional theories. \\
(I) First, following \cite{2dgauge}, we consider the action
\eq
S = \int_\MM d^2 x \Big( -\frac{\l^2}{2} \vf^a \vf^a - i \frac{1}{2}
\ve^\mn \vf^a F^a_\mn \Big),
\eqn{YM}
where $\l$ plays the role of the coupling constant.
After integrating over the field $\vf^a$ in the partition function
\eq
\ZZ = \int D\vf^a DA^a_\m e^{- S}, 
\eqn{partition}
and using the Gaussian integral identity
\eq
\int^{+ \infty}_{- \infty} dx \frac{1}{\sqrt{2 \pi}} ~ exp(
- \frac{\l^2}{2} x^2 - i x y) ~ = ~ exp(-\frac{y^2}{2\l^2}),
\eqn{gauss}
one can easily see that the action (\ref{YM}) gives rise to the same
partition function as the $2D$ Yang--Mills action. Moreover, for 
vanishing coupling constant $\l \to 0$ the action (\ref{YM}) reduces to 
\eq
\int_\MM \frac{-i}{2} \ve^\mn \vf^a F^a_\mn.
\eqn{vf-f}
Hence, in the limit where $(\l \to 0)$ the $2D$ Yang--Mills action and 
(\ref{vf-f}) lead to the same partition function $\ZZ$. \\
(II) Second we consider the $3D$ topological Chern--Simons theory
\eq
\S_{C.S} = - \frac{1}{2} \int d^3 x \ve^{\a\b\g} (A_\a^a \pa_\b A_\g^a +
            \frac{1}{3} f^{abc} A_\a^a A_\b^b A_\g^c),
\eqn{cs}
where the indices $\a, \b$ and $\g$ are the $3D$ space--time indices. 
Now if we compactify the Chern-- Simons model (\ref{cs}) on a circle, 
we get a two dimensional theory described exactly by an action of the 
same form as (\ref{inv}) where the field $\vf^a$ is nothing but 
the third component
of the $3D$ gauge field $A^a_\a$ in the compactified direction. \\
(III) The third possibility we mention is the relation of (\ref{inv})
to $2D$ gravity. Here we follow Chamseddine and Wyler
\cite{2dgravity} where the following action
was proposed as a model for $2D$ gravity
\eq
\S_G = \frac{1}{2} \int d^2 x \ve^\mn \vf^A F^A_\mn.
\eqn{sw}
In this case
\eq
F^A_\mn = \pam e^A_\n - \pan e^A_\m - \e^A_{~BC} e^B_\m e^C_\n,
\eqn{newf}
$A, B$ and $C$ take the values $(0,~1,~2)$ and the generators $\t^A$
of the group $SO(1,2)$ give rise to the following algebra
\eq
\lbrack \t^A, ~ \t^B \rbrack = -\e^{ABC} \t^C .
\eqn{nie}
To explicitly see the connection between a
$2D$ gravity model and the action (\ref{sw}) one let the gauge field
$e^A_\m$ decompose in the zweibein field $e^a_\m$ and the spin connection
$e^2_\m = \o_\m$. After some computations (for details, see 
\cite{2dgravity})
one gets from (\ref{sw}) the action
\eq
I_G = \int d^2 x \sqrt{h} \vf (R + \L),
\eqn{tj}
$h$ is the determinant of the metric $h_\mn = e^a_\m e^b_\n \eta_{ab}$
and $R$ is the corresponding Ricci scalar. $\L$ stands for the  
cosmological constant. At this level one recognizes that beginning with
(\ref{sw}) one could derive the Jackiw--Teitelboim model for $2D$
gravity, given by (\ref{tj}). \\    
%%%%%%%%%%%%%%%%%%%%%%%%%%%%%%%%%%%%%%%%%%%%%%%%%%%%%%%%%%%%%%
Above, we have seen some connections between different field models
and their relation to the action (\ref{inv}), which seems to be the
`meeting point' of different $2D$ theories. Motivated by these facts,
from now on, we will concentrate on the analysis of
the action (\ref{inv}), which is invariant under the gauge symmetry
\eq
\ba{rcl}
\d_g A_\m^a &=& - (\pam \th^a + f^{abc} A_\m^b \th^c) 
\equiv - (D_\m \th)^a, \\
\d_g \vf^a &=& f^{abc} \th^b \vf^c,
\ea
\eqn{;dfsg}
where $\th^a$ is the gauge parameter.
To fix this gauge freedom we choose a Landau--type gauge such that
the gauge fixing part of the action, in which
the metric describing the manifold appears explicitly, writes down as
\begin{equation}
\label{gf}
\Sigma_{gf} = - s \int_{\cal M} d^{2} x \left(
               \sqrt{g} 
               g^{\mu\nu} \partial_{\nu}
               \bar{c}^{a}  A^{a}_{\mu} \right).
\end{equation}
The gauge fixed action $(\Sigma_{inv} + \Sigma_{gf})$ is invariant
under the BRS \cite{brs} symmetry:
\begin{equation}
\label{brsxx}
\begin{array}{ll}
s A^{a}_{\mu} = - ( D_{\mu} c )^{a},                   & \\
s \varphi^{a} = f^{abc} c^{b} \varphi^{c},             & \\
s c^{a} = \frac{1}{2} f^{abc} c^{b} c^{c},             & \\
s \bar{c}^{a} = b^{a} , &     s b^{a} = 0,                \\
s^{2} = 0.                                                \\
\end {array}
\end{equation}
$ c^{a} $ denotes the Faddeev-Popov ghost field of ghost number +1, 
$ \bar{c}^{a}$ is the antighost field of ghost number -1 and 
$ b^{a} $ is the Lagrange multiplier of ghost number 0 enforcing
the Landau gauge condition.\\
As for the other topological gauge field models \cite{lp},\cite{zer},
\cite{mwo}, \cite{ulli}, the metric
$g_{\mu\nu}$ is only present in the gauge fixing part of the 
action which is nothing else but an exact BRS variation. This fact
implies that, here, one can also extend the BRS symmetry \cite{ps} by 
letting the operator s acting on the background metric as:
\begin{equation}
\label{ext.brs}
\begin{array}{ll}
s g_{\mu\nu} = \hat{g}_{\mu\nu}  ~&~  s \hat{g}_{\mu\nu} = 0.
\end{array}
\end{equation}
Thus the metric is just a gauge parameter, of which the 
physical observables are independent.
Clearly, $\hat g_{\m\n}$ is a symmetric tensor of ghost number one.\\
In addition to the BRS symmetry, the gauge fixed action 
$(\S_{inv} + \S_{gf})$ is also invariant under
a local supersymmetry--like transformations (called superdiffeomorphisms
in \cite{lp}, \cite{zer}, \cite{mwo}) and whose parameter we denote by 
$\x^\m$, which has ghost number $+2$.
\eq
\ba{rcl}
\d_{(\x)} A^a_\m &=& 0, \\
\d_{(\x)} \vf^a &=& - \ve_\mn \x^\m \sqrt{g} g^{\n\s} \pa_\s \bar c^a, \\
\d_{(\x)} c^a &=& - \x^\m A^a_\m, \\
\d_{(\x)} \bar c^a &=& 0, \\
\d_{(\x)} b^a &=& \LL_\x \bar c^a, \\
\d_{(\x)} \hat g_\mn &=& \LL_\x g_\mn, \\
\d_{(\x)} g_\mn &=& 0,
\ea
\eqn{mnpq}
where ${\cal L}_{\x}$ denotes the Lie derivative. When we anticommute 
the BRS operator $s$ with $\d_{(\x)}$ we get the on--shell algebra
\eq
\left\{ s , \d_{(\x)} \right\} = \LL_\x + ~equations ~of ~ motion.
\eqn{onshell}
On the other hand, in the context of perturbation theory, the model  
has an infrared problem. Indeed, in the flat space--time limit 
the propagator $\langle c^a \bar c^a \rangle$ is logarithmic divergent 
in the infrared limit. To regularize $\langle c^a \bar c^a \rangle$ one 
has to introduce a mass $m$ \cite{blasi} such that 
\eq
\langle c^a \bar c^a \rangle = \frac{1}{k^2 + m^2}~.
\eqn{prop}
As remarked by the authors of \cite{blasi}, the physical observables
are defined in the limit of vanishing mass. However, as long as $m$
is not zero, the physical quantities may depend on it. A similar
situation was already analyzed in the literature,
where it was conjectured \cite{elitzur} and then shown \cite{bbbcd}
that (in the context of a $2D$ nonlinear sigma model) local observables 
are well defined in the vanishing mass limit.
For topological field theories, however, we have nonlocal observables.
In the spirit of \cite{elitzur}, the authors of \cite{blasi}  
have extended the conjecture of Elitzur \cite{elitzur} to include 
nonlocal observables, too.\\
In this paper we will not worry about such questions but concentrate 
our effort in proving the perturbative finiteness of the model.
Thus, in our case (in the curved space--time) we also introduce a  
mass term $m^2$ which would regularize the propagator 
$\langle c^a \bar c^a \rangle$. Then of course the action gets 
modified\footnote{ We also want to preserve the algebraic structure 
(\ref{onshell}), which we will promote to the functional level
(see below).}
by adding the integrated local polynomial $\S_m$
\eqac
\S_m &=& s \displaystyle{\int}_\MM d^2 x \Big( 
         \t^\m_4 \bar c^a A^a_\m - \sqrt{g} 
         \t_2 \bar c^a c^a \Big) \non \\ 
     &=& \displaystyle{\int}_\MM d^2 x \bigg( 
         \t^\m_3 \bar c^a A^a_\m + 
         \t^\m_4 b^a A^a_\m +
         \t^\m_4 \bar c^a (D_\m c)^a - \frac{1}{2} \sqrt{g}
         g^\mn \hat g_\mn \t_2 \bar c^a c^a + \non \\
     &+& \sqrt{g} (\t_1 + m^2) \bar c^a
         c^a + \sqrt{g} \t_2 b^a c^a - 
         \frac{1}{2} \sqrt{g} \t_2 f^{abc} \bar c^a c^b c^c \bigg)
\eqacn{actionm}
such that 
\begin{equation}
\label{brsm}
\begin{array}{rcl}
s \t_2 &=& - ( \t_1 + m^2 ),   \\   
s \t_1 &= & 0  \\
s \t_4^\m &=& \t^\m_3, \\
s \t_3^\m &=& 0.
\end {array}
\end{equation}
The external sources $\t_1$ and $\t_2$ are scalars  
and $\t^\m_3$ and $\t^\m_4$ are contravariant vector densities
of weight one. 
By just looking at the transformations (\ref{brsm}) we observe that the 
BRS operator $s$ is still nilpotent $s^2=0$ and then $\S_m$ is by 
construction BRS invariant.\\
Another important remark is that the metric $g_{\m\n}$ does still appear 
in the action only through
the BRS exact term $(\S_{gf}+ \S_m)$. Thus the physical observables still
do not 
depend on $g_{\m\n}$, in turn this fact enables us to maintain the BRS
transformation of the metric (\ref{ext.brs}).
In table 1 we give the dimensions, the ghost numbers as well as the 
weights of the above introduced external sources.
\begin{table}[h]
\begin{center}
\begin{tabular}{|c|c|c|c|c|}\hline
      &$\t_1$ &$\t_2$ &$\t^\m_3$ &$\t^\m_4$  \\ \hline
dim   &2  &2  &1  &1  \\ \hline
$\Phi\Pi$  &0  &-1  &1  &0   \\ \hline
Weight  &0 &0 &1 &1  \\ \hline
\end{tabular}
\caption{Dimensions, ghost numbers and weights of the new external 
sources.}
\end{center}
\end{table}
\par

%\newpage
In order to write down the Slavnov identity we couple the nonlinear 
BRS transformations in (\ref{brsxx}) to external sources \cite{bec}, 
which leads to 
the external part of the action $ \Sigma_{ext} $. Hence, the 
total action takes the form:
\eq 
\S = \S_{inv}+\S_{gf}+\S_m+ \S_{ext}.
\eqn{taction}
where,
\begin{equation}
\label{ext}
\Sigma_{ext} =  \int_{\cal M} d^{2} x
               [ \Omega^{a\mu} ( s A^{a}_{\mu} ) +
                 L^{a} ( s c^{a} ) +
                 \rho^{a} ( s \varphi^{a} ) ]
\end{equation}
with $\Omega^{a\mu}$ a contravariant vector
density of weight +1, $L^{a}$ and $ \rho^{a}$ are both
scalar densities of weight +1. In table 2 we give the dimensions, 
the Faddeev-Popov charges and the weights of the different 
fields\footnote{ The vector $\ve^\m$ is the parameter of the 
diffeomorphism transformations (see below).}:
\begin{table}[h]
\begin{center}
\begin{tabular}{|c|c|c|c|c|c|c|c|c|c|c|c|c|}\hline
      &$A^{a}_{\mu}$ &$\varphi^{a}$ &$c^{a}$ &$\bar{c}^{a}$ 
      &$b^{a}$ &$\Omega^{a\mu}$ &$L^{a}$ &$\rho^{a}$ &$g_{\mu\nu}$ 
      &$\hat{g}_{\mu\nu}$ &$\ve^\m$ 
      &$\xi^{\mu}$ \\ \hline
dim   &1  &0  &0  &0  &0  &1  &2  &2  &0  &0  &-1  &-1  
      \\ \hline
$\Phi\Pi$  &0  &0  &1  &-1  &0  &-1  &-2  &-1  &0  &1  
           &1  &2  \\ \hline
Weight  &0 &0 &0 &0 &0 &1 &1 &1 &0 &0 &0 &0 \\ \hline
\end{tabular}
\caption{Dimensions, ghost numbers and weights of the fields.}
\end{center}
\end{table}
\par

Now we are ready to write down the Slavnov identity 
corresponding to the BRS invariance of the total action
(\ref{taction}) at the functional level:
\eqac
\SS (\S)  &=&  \displaystyle{\int}_{\cal M} d^{2}x \bigg(
                    \frac{\d \S}{\d \O^{a\m}}
                    \frac{\d \S}{\d A^a_\m} +
                    \frac{\d \S}{\d \r^a}
                    \frac{\d \S}{\d \vf^a} +
                    \frac{\d \S}{\d L^a}
                    \frac{\d \S}{\d c^a} +
                    b^{a}
                    \frac{\d \S}{\d \bar c^a} +
                    \hat g_{\m\n}
                    \frac{\d \S}{\d g_{\m\n}} +
                    \non \\
             & + &  \t_3^\m \dfrac{\d \S}{\d \t_4^\m}
                    - (\t_1 +m^2) \dfrac{\d \S}{\d \t_2}
                    \bigg) = 0.
\eqacn{slav.taylor1}
From the above Slavnov identity we get 
the linearized Slavnov operator:
\newpage
\begin{eqnarray}
\label{sla.op}
{\cal S}_\Sigma & = & \int_{\cal M} d^{2}x \left(
                    \frac{\delta \Sigma}{\delta \Omega^{a\mu}}
                    \frac{\delta }{\delta A^{a}_{\mu}} +
                    \frac{\delta \Sigma}{\delta A^{a}_{\mu}}
                    \frac{\delta }{\delta \Omega^{a\mu}} +
                    \frac{\delta \Sigma}{\delta \rho^{a}}
                    \frac{\delta }{\delta \varphi^{a}} +
                    \frac{\delta \Sigma}{\delta \varphi^{a}}
                    \frac{\delta }{\delta \rho^{a}} +
                    \frac{\delta \Sigma}{\delta L^{a}}
                    \frac{\delta }{\delta c^{a}} +
                    \right.\nonumber \\
             & + &  \left.\frac{\delta \Sigma}{\delta c^{a}}
                    \frac{\delta }{\delta L^{a}} +
                    b^{a}
                    \frac{\delta }{\delta \bar{c}^{a}} +
                    \hat{g}_{\mu\nu}
                    \frac{\delta }{\delta g_{\mu\nu}} +
                    \t_3^\m \frac{\d}{\d \t^\m_4} -
                    (\t_1 + m^2) \frac{\d}{\d \t_2}
                    \right). 
\eea
%Here $\S$ stands for the total action,
%\eq 
%\S = \S_{inv}+\S_{gf}+\S_m+ \S_{ext}.
%\eqn{taction}
In the presence of the external sources, the Ward operator corresponding
to the local supersymmetry--like transformations writes down as
\eqac
%\label{sdiff.op1}
{\cal W}^{S}_{(\xi)} &=& \displaystyle{\int}_{\cal M} d^{2}x \bigg(
                       \varepsilon_{\mu\nu} \xi^{\mu} \rho^{a}
                       \frac{\delta }{\delta A^{a}_{\nu}} -
                       \varepsilon_{\mu\nu} \xi^{\mu} (
                       \Omega^{a\nu} + 
                       \sqrt{g} g^{\nu\sigma} \partial_{\sigma}
                       \bar{c}^{a} - \t^\n_4 \bar c^a)
                       \frac{\delta }{\delta \varphi^{a}} -
                       \non \\
                 &-&   \xi^{\mu} A^{a}_{\mu}
                       \dfrac{\delta }{\delta c^{a}} +
                       {\cal L}_{(\xi)} \bar{c}^{a}
                       \dfrac{\delta }{\delta b^{a}} -
                       L^{a} \xi^{\mu}
                       \dfrac{\delta }{\delta \Omega^{a\mu}} +
                       {\cal L}_{(\xi)} g_{\mu\nu}
                       \dfrac{\delta }{\delta \hat{g}_{\mu\nu}} -
                       \LL_{(\x)} \t_2 \dfrac{\d}{\d \t_1} +
                       \non \\
                  &+&  \Big( \LL_{(\x)} \t^\n_4 - 
                       \x^\n \sqrt{g}(\t_1 + m^2)
                      + \x^\m (s\sqrt{g}) \t_2
                      \Big) \dfrac{\d}{\d \t^\n_3} -
                      \x^\m \sqrt{g} \t_2 \dfrac{\d}{\d \t^\m_4}
                      \bigg),
\eqacn{sdiff.op1}
and when it acts on the total action (\ref{taction}) we get the linear 
breaking
\eq
\WW_{(\x)} \S = \D_{(\x)},
\eqn{111}
where,
\eq
\ba{rcl}
\D_{(\x)} &=& \displaystyle{\int}_\MM ~d^2x \bigg(
              - \O^{a\m} \LL_{(\x)} A_\m^a -
           \r^a \LL_{(\x)} \vf^a + L^a \LL_{(\x)} c^a +
          \non \\
&+&  \displaystyle{\ve}_{\m\l} \x^\l \r^a s(\sqrt{g} g^{\m\n} 
      \pan \bar c^a) - \ve_{\m\n} \x^\n \r^a s (\t^\m_4 \bar c^a) 
\bigg).
\ea
\eqn{br}
At this level we see that the Ward identity of the local susy--like 
symmetry is linearly broken. This is not the case for the topological
Yang-Mills model \cite{zer}, the three dimensional BF model \cite {mwo}, 
and the Chern-Simons model \cite{lp} considered in a curved manifold, 
where one had to do with a hard breaking.\\
By construction $\S$ is also invariant under the diffeomorphism 
transformations
\eq
\WW^D_{(\ve)} \S = \int_\MM d^2 x \sum_f \LL_\ve f \frac{\d \S}{\d f} = 0.
\eqn{diff.inv}
The letter f stands for all the fields describing the model
under investigation, whereas $\ve^\m$ is the parameter
of the diffeomorphism transformations and ${\cal L}_{\ve}$ 
denotes the corresponding Lie derivative.
Furthermore, the action $\S$ obeys three constraints: \\
(i) the gauge condition
\eq
\frac{\d \S}{\d b^a} = \pan (\sqrt{g} g^{\m\n} A_\m^a) + \sqrt{g} \t_2 c^a
                       + \t^\m_4 A_\m^a,
\eqn{mgc}
(ii) the antighost equation
\eq
\frac{\d \S}{\d \bar c^a} + \pan (\sqrt{g} g^{\m\n} \frac{\d \S}{\d \O^{a\m}})
+ \t^\m_4 \frac{\d \S}{\d \O^{a\m}} - \sqrt{g} \t_2 \frac{\d \S}{\d L^a} =
\pan \Big( s(\sqrt{g} g^{\m\n}) A_\m^a \Big) + \sqrt{g} 
(\t_1 + m^2) c^a - 
\t^\m_3 A_\m^a,
\eqn{mage}
(iii) and the ghost equation
\eq
\int_\MM d^2x \left( \frac{\d S}{\d c^a} + f^{abc} \bar c^b
\frac{\d S}{\d b^c}
\right) = \int d^2x \Big( f^{abc} ( \O^{b\m} A_\m^c - L^b c^c
+ \r^b \vf^c) - \sqrt{g}(\t_1 + m^2) \bar c^a - \sqrt{g} \t_2 b^a \Big).
\eqn{mge}
To obtain the ghost equation, one has simply to integrate over the 
space--time the quantity $(\dfrac{\d \S}{\d c^a})$ and then use the
gauge condition (\ref{mgc}).\\
We end this section by displaying the algebraic structure of the model.
First, consider an arbitrary local functional $\G$ depending on 
the fields $(A_\m^a$, $\vf^a$, $c^a$, $\bar c^a$, $b^a$, $\O^{a\m}$, 
$L^a$, $\r^a$, $g_{\m\n}$, $\hat g_{\m\n}$, $\t_1$, $\t_2$, $\t^\m_3$, 
$\t^\m_4)$, one can derive the following nonlinear algebra
\eq
\ba{rcl}
\SS_\G \SS (\G) &=& 0, \\
\SS_\G (\WW^D_{(\ve)}) + \WW^D_{(\ve)} \SS (\G) &=& 0, \\
\left\{\WW^D_{(\ve)}, \WW^D_{(\ve^\prime)} \right\} \G &=&
- \WW^D_{([\ve, \ve^\prime])} \G, \\
\SS_\G (\WW^S_{(\x)} \G - \D_{(\x)}) + \WW^S_{(\x)} \SS (\G) &=& 
\WW^D_{(\x)} \G, \\
\left\{ \WW^D_{(\ve)}, \WW^S_{(\x)} \right\} \G &=& 
\WW^S_{([\x, \ve])} \G, \\
\left\{ \WW^S_{(\x)}, \WW^S_{(\x^\prime)} \right\} \G &=& 0.
\ea
\eqn{alg}  
Second, if the functional $\S$ obeys the Slavnov identity as well as 
the two Ward identities for diffeomorphisms and local susy--like 
symmetry, then we get the linear off--shell algebra  
\eq
\ba{rcl}
\left\{ \SS_S, \SS_\S \right\} & = & 0,  \\
\left\{ \SS_S, \WW^{D}_{(\ve)} \right\} & = & 0, \\
\left\{ \WW^{D}_{(\ve)},\WW^{D}_{(\ve^{\prime})}
\right\} & = & - \WW^{D}_{([\ve, \ve^\prime])}, \\
\left\{ \SS_\S, \WW^{S}_{(\xi)} \right\}
& = & \WW^{D}_{(\xi)}, \\
\left\{ \WW^{S}_{(\xi)}, \WW^{D}_{(\ve)} \right\}
& = & \WW^{S}_{([\xi,\ve])}, \\
\left\{ \WW^{S}_{(\xi)}, \WW^{S}_{(\xi^{\prime})} \right\}
& = & 0.
\ea
\eqn{lalg} 
Here, we have used the following notation
\eq
\ba{rcl}
\lbrack \ve,~ \ve^\prime \rbrack^\m & = & 
\ve^\l \pal \ve^{\prime \m} + \ve^{\prime \l} \pal \ve^\m, \\
\lbrack \xi,~\ve \rbrack^\m & = & \xi^\l \pal \ve^\m 
- \ve^\l \pal \xi^\m.
\ea
\eqn{tym1}  
Where $[~,~]$ stands for the graded Lie bracket. Furthermore, for 
reasons which will be clear in the next section, we have 
attributed ghost number one to $\ve^\m$, the vector parameter of the
diffeomorphism transformations.\\ 
We conclude this section by remarking that the mass $m$, used to 
regularize the infrared divergent propagator, as in (\ref{prop}) could 
destroy the algebraic structure (\ref{onshell}) of the model at the 
off-shell level (\ref{lalg}). 
To maintain this structure, in the presence of a curved 
background, the price to pay was the introduction of four new external 
sources \cite{blasi} $\t_1$, $\t_2$, $\t_3^\m$ and $\t_4^\m$, which 
appear in the action through the metric dependent and BRS--exact 
expression $\S_m$.  

\section{Cohomology analysis}

In this section we will look for all possible quantum corrections for the 
model. Indeed,
the construction of the most general counterterm can be done as follows, first
we add a perturbation $\D$ to the total action $\S$ such that the perturbed 
action $\S' = \S + \D $ fullfils the Slavnov identity (\ref{slav.taylor1}), 
and the two Ward identities (\ref{111}), (\ref{diff.inv}) as well 
as the identities (\ref{mgc}), (\ref{mage}) and (\ref{mge}). 
%conside the purturbe$\S' = \S + \D $, 
Therefore $\D$ must obey the constraints:
\eq
\frac{\d \D}{\d b^a} = 0,
\eqn{cst1}
\eq
\frac{\d \D}{\d \bar c^a} + \pan (\sqrt{g} g^{\m\n} \frac{\d \D}{\d \O^{a\m}})
+ \t^\m_4 \frac{\d \D}{\d \O^{a\m}} -  \sqrt{g} \t_2 \frac{\d \D}{\d L^a} = 0,
\eqn{cst2}
\eq
\int_\MM d^2 x \frac{\d \D}{\d c^a} = 0,
\eqn{cst3}
\eq
\SS_\S \D = 0,
\eqn{cst4}
\eq
\WW^D_{(\ve)} \D = 0,
\eqn{cst5}
\eq
\WW^S_{(\x)} \D = 0.
\eqn{cst6}
$\D$ is an integrated local polynomial of dimension, weight and ghost 
number zero.
From equation (\ref{cst1}) we immediately see  that $\D$ does not depend on the
Lagrange multiplier field $b^a$. On the other hand, from (\ref{cst2}) we
deduce that the integrated polynomial $\D$ can depend on the fields 
$\O^{a\m}$, $\bar c^a$ and $L^a$ only through the combinations
\eq
\ba{rcl}
\Ot^{a\m} &=& \O^{a\m} + \sqrt{g} g^{\m\n} \pan \bar c^a - \t^\m_4
\bar c^a , \\
\tilde L^a &=& L^a - \sqrt{g} \t_2 \bar c^a .
\ea
\eqn{comb}
Concerning the three equations (\ref{cst4}) -- (\ref{cst6}), we put them in
a single equation 
\eq
\d \D = 0
\eqn{co1}
such that the operator $\d$ is of the form 
\eq
\d = \SS_\S + \WW^D_{(\ve)} + \WW^S_{(\x)} + \int_\MM d^2 x 
( \LL_{\ve} \x^\m) \frac{\d }{\d \x^\m} + 
\int_\MM d^2 x \left( \frac{1}{2} \LL_{\ve} \ve^\m
- \x^\m \right) \frac{\d }{\d \ve^\m}.
\eqn{delta}
An easy check is to show its nilpotency,
\eq
\d^2 = 0,
\eqn{nil}
so that (\ref{co1}) is a cohomology problem possessing two possible 
solutions. Indeed, (\ref{co1}) possesses solutions of the form 
$\delta = \delta \hat \Delta $. These are called trivial solutions 
because the nilpotency of $\delta$ immediately implies that any 
expression of the form $\delta \hat \Delta$ is automatically a 
solution of (\ref{co1}). In what follows we will call 
{\it cohomology of $\delta$} the space of all
solutions of (\ref{co1}) modulo trivial solutions.\\
The first step in solving the cohomology problem (\ref{co1}) 
is to introduce
a filtering operator $\NN$ and assigne to each field (including $\ve^\m$ 
and $\x^\m$) homogeneity degree $1$. 
\eq
\NN = \int_\MM d^2 x \sum_f f \frac{\d }{\d f}  + \int_\MM d^2 x
\left( \ve^\m \frac{\d }{\d \ve^\m} + \x^\m \frac{\d }{\d \x^\m}
\right).
\eqn{NNN}
The operator ${\cal N}$ induces a decomposition of $\delta$
\begin{equation}
\label{dec}
\delta = \delta_0 + \delta_1 + ... + \delta_N,
\end{equation}
as well as a decomposition of $\Delta$ 
\begin{equation}
\Delta = \sum_{n\geq0} \Delta_n,
\end{equation}
where the index $n$ refers to the corresponding homogeneity degree.
The operator $\delta_0$ in (\ref{dec}) has the property that it does
not increase the homogeneity degree when it acts on a field polynomial.
On the other hand, due to the nilpotency of $\delta$ we also have
\begin{equation}
\label{xxx}
\delta_0^2 = \lbrace \delta_0 , \ \delta_1 \rbrace = 0,
\end{equation}
and more generally 
\begin{equation}
\sum_{i=0}^{k} \delta_i \delta_{k-i} = 0 ; \ \ \ \ \ k \leq N.
\end{equation}
An obvious identity which follows from (\ref{dec}) and 
$\delta \Delta = 0$ reads as
\begin{equation}
\label{co2}
\delta_0 \Delta = 0,
\end{equation}
Due to the nilpotency of $\delta_0$ (\ref{xxx}), the above equation 
(\ref{co2}) 
defines a new cohomology problem. 
An interesting result is the following theorem\footnote{ An immediate
corollary of the theorem is as follows: if the cohomology 
of $\delta_0$ is empty (trivial), then that of $\delta$ is also
empty.}\\
\newpage
{\sl Theorem:} \\
{\it The cohomology of the operator $\delta$ is isomorphic to a 
subspace of the cohomology of the operator $\delta_0$.} \\
More concretely, for the $2D$ model under investigation, we have
\eqac
\d_0 &=& \displaystyle{\int}_\MM \left( d c^a \frac{\d }{\d A^a} + 
         d A^a \frac{\d }
         {\d \hat \r^a} + d \vf^a \frac{\d }{\d \hat \O^a} + 
         d \hat \O^a \frac{\d }{\d \hat L^a} \right) + \non \\
    &+&  \displaystyle{\int}_\MM d^2 x \bigg( \hat g_\mn 
         \frac{\d }{\d g_\mn} -
         \t_1  \frac{\d }{\d \t_2} + 
         \t^\m_3 \frac{\d }{\d \t^\m_4} -
         \x^\m \frac{\d }{\d \ve^\m} \bigg)
\eqacn{delta0}
The first part of the expression of $\d_0$ is given in terms of forms
where,
\eq
\ba{rcl}
A^a &=& A_\m^a \dxm, \\
\hat \O^a &=& \ve_\mn \Ot^{a\m} \dxn, \\
\hat L^a &=& \frac{1}{2} \ve_\mn \tilde L^a \dxm \dxn, \\
\hat \r^a &=& \frac{1}{2} \ve_\mn \r^a \dxm \dxn, \\
\ea
\eqn{forms}
and $d$ is the exterior derivative $d = \dxm \pam$.\\
one can easily see from (\ref{delta0}) that the following couples of 
fields $(g_\mn, \hat g_\mn)$, $(-\t_1, \t_2)$, $(\t^\m_3, \t^\m_4)$ and
$(\ve^\m, -\x^\m)$ appear in $\d_0$--doublets, and then are not present 
in the cohomology \cite{brandt}.
Now, let us solve $\d_0 \D = 0$. The local integrated polynomial $\D$
can be written as
\eq
\D = \int_\MM \o^0_2
\eqn{lkj}
where $\o^p_q$ is a polynomial of ghost number $p$ and form degree $q$.
By letting the operator $\d_0$ acting on (\ref{lkj}) and taking 
into account (\ref{co2}) we get, after using Stock's theorem
\eq
\d_0 \o^0_2 + d \o^1_1 = 0
\eqn{des1}
Now, by applying once again $\d_0$ on equation (\ref{des1}) and using the
algebraic Poincare lemma\footnote{Roughly speaking, the algebraic Poincare 
lemma states that, in the space of forms depending on the fields and their 
derivatives, the cohomology of the exterior derivative $d$ is trivial.
For the exact formulation and proof of the lemma see \cite{brandt}.} 
\cite{brandt} and the facts that 
$\d_0^2 = 0$ and 
$\left\{\d_0 , d \right\}= 0$ we get the descent equations
\eq
\ba{rcl}
\d_0 \o^0_2 + d \o^1_1 &=& 0, \\
\d_0 \o^1_1 + d \o^2_0 &=& 0, \\
\d_0 \o^2_0 &=& 0.
\ea
\eqn{des}
It is clear that the solution\footnote{where $\vf = \vf^a T^a$ and 
$c = c^a T^a$. $Tr$ is the trace defined
by $Tr (T^a T^b)= \d^{ab}$ and $T^a$ are the generators of the gauge group.} 
of the last equation in (\ref{des}) is 
\eq
\o^2_0 = \sum_{n=0}^\infty \sum_{m=0}^\infty \a_{n,m}
Tr (\vf^n c \vf^m c).
\eqn{dggtr}
where $\a_{n,m}$ are constant coefficients. The condition that 
the solution of (\ref{co1}) must be invariant under the whole operator
$\d$ implies that $\a_{n,m}=0$ unless $m=0$. In this case we define 
$\a_{n,0} \equiv \a_n$. In turn, $\o^2_0$ takes the form
\eq
\o^2_0 = \sum_{n=1}^\infty  \a_n Tr (\vf^n c c).
\eqn{bbla}
For $n=0$, of course, one has $\o^2_0 = Tr (c^2) = c^a c^a = 0$.
The expression (\ref{bbla}) leads to the nontrivial counterterm
\eq
\ba{rcl}
\D_c &=& \displaystyle{\sum}_{n=1}^{\infty} \a_n Tr \displaystyle{\int}_\MM 
\bigg( \sum^{n-1}_{i=0}
\sum^{i-1}_{j=0} \vf^j \hat \O \vf^{i-j-1}  \hat \O
\vf^{(n-i-1)}c^2 + \sum^{n-1}_{i=0} \vf^i \hat L \vf^{(n-i-1)} c^2
+ \non \\
&+& \displaystyle{\sum}^{n-1}_{i=0} \vf^i \hat \O  \sum^{n-i-2}_{j=0} 
\vf^j \hat \O
\vf^{(n-i-j-2)} c^2 + 2 \sum^{n-1}_{i=0} \vf^i \hat \O \vf^{(n-i-1)} 
\lbrace A, c \rbrace 
+ \non \\
&+& 2 \vf^n \lbrace \hat \r, c \rbrace  + 2 \vf^n A^2  
\bigg),
\ea
\eqn{ntriv}
which is invariant under the whole operator $\d$ defined in (\ref{delta}).
Furthermore, the trivial solution $\d \hat \D$ of (\ref{co1}) 
is given by
\eq
\ba{rcl}
\d \hat \D &=& \d \displaystyle{\int}_\MM d^2 x \Big\lbrace  
\sqrt{g} \t_2 
f_1(\vf)  + \dfrac{1}{\sqrt{g}} 
g_\mn \t_4^\m \sum^{\infty}_{n=1} \b_{1,n} 
Tr (\Ot^\n \vf^n) + \non \\
&+&
\ve_\mn \t_4^\m \sum^{\infty}_{n=1} \b_{2,n}
Tr (\Ot^\n \vf^n) + \r^a \vf^a [\k_1 + f_2(\vf) ] + 
\tilde L^a c^a [\k_2 + f_3(\vf) ] + \non \\
&+&
\ve_\mn \sqrt{g} g^{\n\r} \Ot^{a\m} A^a_\r [\k_3 + f_4(\vf) ] +
\Ot^{a\m} A^a_\m [\k_4 + f_5(\vf) ] +  \non \\
&+& 
\sum^{\infty}_{n=1} \b_{3,n} Tr (\pam \Ot^\m \vf^n) + 
\ve_\mn \sqrt{g}
g^{\r\n} \sum^{\infty}_{n=1} \b_{4,n} Tr ( \Ot^\m \pa_\r \vf^n ) 
+ \non \\
&+&
\sum^{\infty}_{n=1} \b_{5,n} Tr ( \pam \Ot^\m \vf^n ) 
+ g^\mn \hat g_\mn \sum^{\infty}_{n=1} \b_{6,n} 
Tr ( \tilde L^a \vf^n )  \Big\rbrace.
\ea
\eqn{triv}
($\k_1, ..., \k_4$) and ($\b_{1,n}, ..., \b_{6,n}$)
%with $(1 \leq i \leq 11)$ 
are constant coefficients and
the functions $f_i$ with $(1 \leq i \leq 5)$ are given by
\eq
f_i (\vf) = \sum_{n=2}^\infty \a_{i,n} Tr \vf^n.
\eqn{fi.def} 
On the other hand, the trivial counterterm (\ref{triv}) would depend on
the transformation parameters $\x^\m$ and $\ve^\m$ which are present
in $\d$ (\ref{delta}). The requirement that (\ref{triv}) must 
be independent of this two transformation parameters reduces 
the trivial counterterm to the simpler form
\eq
\ba{rcl}
\d \hat \D &=& \SS_\S \displaystyle{\int}_\MM d^2 x ~\k (\r^a \vf^a + 
\tilde L^a c^a - 
\Ot^{a\m} A^a_\m ) , \non \\
&\equiv& \SS_\S \bar \D,
\ea
\eqn{triv2}
with $\k$ a constant.
So, the possible deformation of the total action $(\ref{taction})$ is 
of the general form
\eq
\D = \D_c + \SS_\S \bar \D.
\eqn{deformation}
In constructing the counterterm (\ref{deformation}) we took into
account all the constraints (\ref{cst1})--(\ref{cst6}) except the
ghost equation (\ref{cst3}). In fact, the expression (\ref{deformation})
violates (\ref{cst3}) unless all the constant coefficients appearing in
(\ref{deformation}) vanish. Therefore the action
(\ref{taction}) admits no deformations and all the quantities present at
the classical level remain the same and receive no corrections.
Furthermore,
if the constraints hold at the quantum level and the symmetries are not
broken, then the complete absence of deformations would imply the
absence of quantum corrections. In this case the theory is said to 
be finite.  

\section{Anomaly analysis}

The last point to be discussed is the possibility of extending the above
analysis to all orders of perturbation theory. This fact is only allowed
when anomalies are absent .\\
The three conditions (\ref{cst1}), (\ref{cst2}) and (\ref{cst3}) can be 
proven to be renormalizable at all orders of perturbation theory by 
using the arguments of \cite{olivier2} and \cite{bps}. 
Concerning the Slavnov identity and the two Ward identities for 
diffeomorphisms and local susy--like symmetry, if there is an anomaly,
then for the generating functional of vertex functions $\G=\S+O(\hbar)$
we must have 
\eq
\d \G = \AA.
\eqn{jg}
Due to the nilpotency of $\d$ we get a new cohomology problem
\eq
\d \AA = 0,
\eqn{anco}
where $\AA$ is an integrated local polynomial of form degree $2$ and
ghost number $1$. 
\eq
\AA = \int_\MM \o^1_2.
\eqn{lskjfg}
Now, using the same strategy as explained in the previous section,
we get the following set of descent equations
\eq
\ba{rcl}
\d_0 \o^1_2 + d \o^2_1 &=& 0, \\
\d_0 \o^2_1 + d \o^3_0 &=& 0, \\
\d_0 \o^3_0 &=& 0.
\ea
\eqn{anom}
The last equation in (\ref{anom}) is solved by 
\eq
\o^3_0 = \sum^\infty_{n=0} \sum^\infty_{m=0} \sum^\infty_{r=0}
\a_{n,m,r} Tr (\vf^n c \vf^m c \vf^r c ).
\eqn{bla2*}
This will yield an expression of $\AA$ which is not invariant under
the whole operator $\d$, $\ie$ $\d \AA \not= 0$ unless $\a_{n,m,r} = 0$ 
for all nonvanishing values of $n$,
$m$, and $r$. This particularly means that we are left with the single
term $\o^3_0 = \a f^{abc} c^a c^b c^c $, which leads to
\eq
\o^1_2 = \a f^{abc} (\hat \r^a c^b c^c + A^a A^b c^c).
\eqn{bla1*}
A quick verification shows that $\o^1_2$ is invariant under the whole
operator $\d$. Hence, the possible anomaly candidate solving 
(\ref{anco}) takes the form 
\eq
\AA = \a \int_\MM f^{abc} (\hat \r^a c^b c^c + A^a A^b c^c ),
\eqn{ano11}
where $\a$ stands for a constant coefficient. But this anomaly candidate
violates the ghost equation (\ref{cst3}), a fact which imposes the 
restriction $\a = 0$. The final result is as follows: the BRS symmetry, the 
diffeomorphisms and the local susy--like symmetry are anomaly free, then
valid at the quantum level. Therefore, the $2D$ model is anomaly free and
ultraviolet as well as infrared finite at all orders of perturbation theory. 
However, as already mentioned in the introduction, our proof of the 
finiteness is only valid in the case of a topologically trivial and 
asymptotically flat manifolds.

\section*{Acknowledgements}

I would like to thank M.W. de Oliveira for useful discussions.
This work is supported in part by the ``Fonds zur F\"orderung 
der Wissenschaftlichen Forschung'' under 
contract grant number P11582--PHY.


\newpage

\begin{thebibliography}{99}
\addcontentsline{toc}{section}{Bibliography}

\bibitem{blau93}M. Blau, G. Thompson, hep-th/9310144.
%
\bibitem{witcs}S.S. Chern and J. Simons, Ann. Math. 
{\bf 99}(1974)48; \\
E. Witten, Comm. Math. Phys. {\bf 121}(1989)351;\\
J. Fr\"ohlich and C. King, \cmp{126}{89}{167}; \\
S. Axelrod and I.M. Singer,
{\it Chern-Simons perturbation theory}, preprint submitted to the
proc. of XXth Conf. on Differential Geometric Methods in Physics,
Baruch College, CUNY (1991).
%
\bibitem{2dgravity}T. Fukuyama and K. Kamimura, Phys. Lett.
{\bf B160}(1985)259; \\
K. Isler and C.A. Trugenberger, Phys. Rev. Lett. 
vol.{\bf 63}(1989)834.\\ 
A.H. Chamseddine and D. Wyler, \np{B340}{90}{595};
Phys. Lett. {\bf B228}(1989)595; \\
D. Cangemi, R. Jackiw and B. Zwiebach, Ann. Phys. (N.Y.)
{\bf 245}(1996)408.
%
\bibitem{2dgauge}E. Witten, Comm. Math. Phys. 
{\bf 141}(1991)153; \\
M. Blau and G. Thompson, Ann. Phys. (N.Y.) {\bf 205}(1991)130; \\
J. Soda, Phys. lett. {\bf B267}(1991)214.
%
\bibitem{blasi}A. Blasi and N. Maggiore, class. Quantum Grav. {\bf 10} 
(1993)37.
%
\bibitem{ms}N. Maggiore and S.P. Sorella, Nucl. Phys. {\bf B377}
(1992)236.
%
\bibitem{mwo}M.W. de Oliveira, M. Schweda and H. Zerrouki, 
Helv. Phys. Acta {\bf 68} 
(1995)73.
%
\bibitem{olivier2} O. Piguet and S.P. Sorella,
{\it Algebraic Renormalization},
Lecture Notes in Physics, Vol. m28, Springer Verlag, 1995.
%
\bibitem{bps}A. Blasi, O. Piguet and S.P. Sorella, Nucl. Phys.
{\bf B356}(1991)154.
%
\bibitem{dix}J.A. Dixon,``Cohomology and Renormalization of Gauge
Theories'', preprints Imperial College, 1977; Comm. Math. Phys. 
{\bf 139}(1991)495.
%
\bibitem{brandt} F. Brandt, N. Dragon and M. Kreuzer,
\pl{B231}{89}{263} \\
\np{B332}{90}{224} \\
\np{B340}{90}{187}
%
\bibitem{sor}S.P. Sorella, Comm. Math. Phys. {\bf 189}(1993)456.
%
\bibitem{lp}C. Lucchesi and O. Piguet, Phys. Lett. {\bf B271}
(1991)350; Nucl. Phys. {\bf B381}(1992)281. 
%
%\bibitem{verlinde} E. Verlinde and H. Verlinde, Nucl. Phys.
% {\bf B348}(1991)457.
%
\bibitem{oli} O. Piguet, {\it On the Role of Vector Supersymmetry
in Topological Field Theory}, preprint UVGA -- DPT 1995/02-880;
hep-th/9502033.
%
\bibitem{brs}C.M. Becchi, A. Rouet and R. Stora, Comm. Math. Phys. 
{\bf 42}(1975)127; Ann. Phys. (N.Y.) {\bf 98}(1976)287.
%
\bibitem{brt}D. Birmingham, M. Rakowski and G. Thompson,
Nucl. Phys. {\bf B329}(1990)83. %
%
\bibitem{tft}D. Birmingham, M. Blau, M. Rakowski and G. Thompson,
Phys. Rep. {\bf 209}(1991)129. 
%
%\bibitem{lp}C. Lucchesi and O. Piguet, Phys. Lett. {\bf B271}
%(1991)350; Nucl. Phys. {\bf B381}(1992)281. 
%
\bibitem{bec}C.M. Becchi,``The Renormalization of Gauge Theories'', 
in the Proceedings of Les Houches Summer School 1983, 
eds. B.S. De Witt and R. Stora, North Holland, Amsterdam, 1984.
%
\bibitem{ps}O. Piguet and K. Sibold, Nucl. Phys. {\bf B253}(1985)517.
%
\bibitem{bbbcd}C.M. Becchi, A. Blasi, G. Bonneau, R. Collina and F.
Delduc, Comm. Math. Phys. {\bf 120}(1988)121.
%
\bibitem{elitzur}S. Elitzur, Nucl. Phys. {\bf B253}(1983)536.
%
\bibitem{pr}O. Piguet and A. Rouet, Phys. Rep. {\bf 76 C}(1981)1.
%
\bibitem{piguet} O. Piguet and K. Sibold,
{\it Renormalized Supersymmetry}, series ``Progress in Physics'',
vol.12 (Birkh\"auser Boston Inc., 1986);
%
\bibitem{zer}H. Zerrouki, Mod. Phys. Lett. {\bf A10} (1995)2253.
%
\bibitem{emery} S. Emery, O. Moritsch, M. Schweda, T. Sommer and 
H. Zerrouki, Helv. Phys. Acta {\bf 68} (1995)167.
%
\bibitem{ulli} U. Feichtinger, O. Moritsch, J. Rant, M. Schweda and 
H. Zerrouki, hep-th/9611070.



\end{thebibliography}
\end{document}